\documentclass[useAMS,usenatbib]{mn2e}
\usepackage{graphicx}

\title[X-ray observations of VY Scl-type nova-like stars]
{X-ray observations of VY Scl-type nova-like binaries in the high and low state}
\author[Zemko P., Orio, M., Mukai, K., Shugarov S.]
{Zemko P.$^{1}$\thanks{E-mail:polina.zemko@studenti.unipd.it}; Orio M.$^{2,3}$\thanks{E-mail:marina.orio@oapd.inaf.it}; Mukai, K.$^{4,5}$\thanks{E-mail:koji.mukai@umbc.edu}; Shugarov S.$^{6,7}$\thanks{E-mail:shugarov@ta3.sk}\\
$^{1}$ Department of Physics and Astronomy, Universit\`a di Padova, vicolo dell' Osservatorio 3, I-35122 Padova, Italy\\
$^{2}$ INAF - Osservatorio di Padova, vicolo dell' Osservatorio 5, I-35122 Padova, Italy\\
$^{3}$ Department of Astronomy, University of Wisconsin, 475 N. Charter Str., Madison, WI 53704, USA\\
$^{4}$ CRESST and X-ray Astrophysics Laboratory, NASA Goddard Space Flight Center, Greenbelt, MD 20771, USA \\
$^{5}$ Department of Physics, University of Maryland, Baltimore County, 1000 Hilltop Circle, Baltimore, MD 21250, USA\\
$^{6}$ Sternberg Astronomical Institute, Moscow University, Universitetsky Ave., 13, Moscow 119992, Russia\\
$^{7}$ Astronomical Institute of the Slovak Academy of Sciences, Tatransk\'{a} Lomnica, 059 60, The Slovak Republic}

\begin{document}

\date{Accepted . Received ; in original form }

\pagerange{\pageref{firstpage}--\pageref{lastpage}} \pubyear{}

\maketitle

\label{firstpage}

\begin{abstract}
Four VY Scl-type nova-like systems were observed in X-rays during both the low and the
 high optical states.
We examined {\sl Chandra}, {\sl ROSAT}, {\sl Swift} and {\sl Suzaku}
 archival observations of BZ Cam, MV Lyr, TT Ari, and V794 Aql.
 The X-ray flux of BZ Cam is higher 
 during the low state, but there is no supersoft X-ray source (SSS) as
 hypothesized in previous articles. No SSS was detected in the 
 low state of the any of the other systems, with the X-ray flux decreasing by a factor
between 2 and 50.
The best fit to the {\sl Swift} X-ray spectra is obtained with
a multi-component model of plasma in collisional ionization
 equilibrium. The high state
high resolution spectra of TT Ari taken with {\sl Chandra} ACIS-S and the HETG
 gratings show a rich emission line spectrum, with prominent lines of in Mg, Si, Ne, and 
 S. The complexity of this spectrum seems to have origin in more than one region,
 or more than one single physical mechanism. While several emission lines are consistent with 
 a cooling flow in an accretion stream, there is at least an additional component.  
 We discuss the origin of this component, which is probably arising in a wind
from the system. We also examine the possibility that the VY Scl systems may be
 intermediate polars, and that while the boundary layer of the accretion disk 
 emits only in the extreme ultraviolet, part of the X-ray flux may be due to magnetically
 driven accretion. 
\end{abstract}


\section{Introduction}
Nova-like (NLs) stars are non-eruptive cataclysmic variables  \citep[CVs, see][]{war95book},
 classified into several subtypes depending on evidence of a strong magnetic field on 
 the white dwarf (WD), and on spectral and photometric characteristics 
\citep[see][]{dhi96vyscl}. Here we will focus on the VY Scl-type nova-likes or 
`anti-dwarf' novae characterized by the presence of 
occasional dips on the light curve, so-called low states, defined by 
\citet{hon04vyscl} as a fading of the optical light by more than 1.5 mag in less than 150 
days. The drop in luminosity can reach 7 mag and may last from weeks to years. 

The large optical and UV luminosity seems to imply that the VY Scl-type nova-like, in 
their longer lasting optically high state, are undergoing mass transfer onto the WD at 
the high rate $\dot{m}>10^{-10}$ M$_{\odot}$ yr$^{-1}$, sustaining an accretion disk
 in a stable hot state in which DN outbursts are suppressed 
 \citep[see e.g. the disk thermal instability model of][]{osa05DImodel}.
The low states have been attributed to a sudden drop of $\dot{m}$ from the secondary, or 
even to a total cessation of mass transfer \citep[see e.g.][]{kin98DI, hes00CVlowstate}. 
The reason of this dramatic decrease of $\dot{m}$ is still unclear. 
The most probable cause may be spots on the surface of the secondary, covering the L1 
point and causing the mass-transfer cut off \citep{liv94CVstarspot}. 
\citet{wu95vyscl} have suggested instead non-equilibrium effects in the irradiated 
atmosphere of the donor.

If the transition from the high to low state occurs because of a decrease in $\dot{m}$, 
the accretion disks in these systems should move from the equilibrium region to the one of
 dwarf novae (DN) instabilities, so we should observe (DN) outbursts with
 recurrence times of 12 to 20 days during the low state, 
caused by thermal--viscous instabilities in the accretion disk \citep{war95book}. 
However, outbursts in the low state of these objects are extremely rare 
\citep[see][for MV Lyr]{pav99mvlyr}.
DN outbursts must be suppressed during the low state despite the low $\dot m$; this can be
 explained by a WD effective temperature high enough (30,000--50,000 K) to irradiate the 
 inner accretion disk and maintain it in the stable `hot' state \citep{sma83DN}, while 
 the incoming mass accretion stream stops
 or decreases \citep{kin97UVdelay, las99DI, ham99uvdelay, lea99vyscl}. 
 The WD in the known DN never reaches this temperature range, but 
high WD effective temperatures have indeed been inferred via spectroscopic
 observations in the {\it UV} and {\it FUV} ranges for the VY Scl objects (see Table 1).
 \cite{ham02vysclmagnetic} suggested instead that the DN outbursts are prevented by the 
 periodic disruption of the disc by a magnetic field of the WD, and in this model the VY 
 Scl would be intermediate polars (IP), in which the WD magnetic field is of the order of 
 10$^6$ Gauss.

\citet{gre99v751cyg} proposed a link between the VY Scl-type stars and super
soft X-ray sources (SSS) based on a {\sl ROSAT} observations of V751 Cyg. They 
found an anti-correlation in the optical and X-ray intensity, and despite the very poor 
spectral resolution of the {\sl ROSAT} HRI, the spectrum appeared to be soft in the low state. 
The authors suggested that quasi-stable thermonuclear burning occurs on the surface of 
the WD in the low state, preventing DN outbursts. In this framework, VY Scl stars
 are key objects in the evolution of interacting WD binaries, in which
 hydrogen burning occurs periodically without outbursts. Accretion goes on at a very high 
 rate without ever triggering a thermonuclear runaway, because of the recurrent 
 interruptions of the high $\dot{m}$ regime. Thus there is a possibility that VY Scl stars
  reach the Chandrasekhar mass and the conditions for type Ia supernovae outbursts.
\citet{gre98vsgeSSS} and \citet{gre01bzcam} suggested also thermonuclear burning in the 
low states of V Sge and BZ Cam. However, these two objects had not been 
 actually observed as SSS; in more recent years an X-ray observation of the VY Scl system 
 V504 Cyg in the low state failed to reveal a luminous SSS \citep{gre10v504cenvyscl}.

Using archival X-ray observations, in this paper we compare high and low state X-ray 
data, and some new {\it UV} data, for four VY Scl-type stars. We seek clues to the complex evolution
 of these systems, and explanations for the changes that take place during the 
 transition from the high to low state. 

\begin{figure}
\includegraphics[width=250pt, height=400pt]{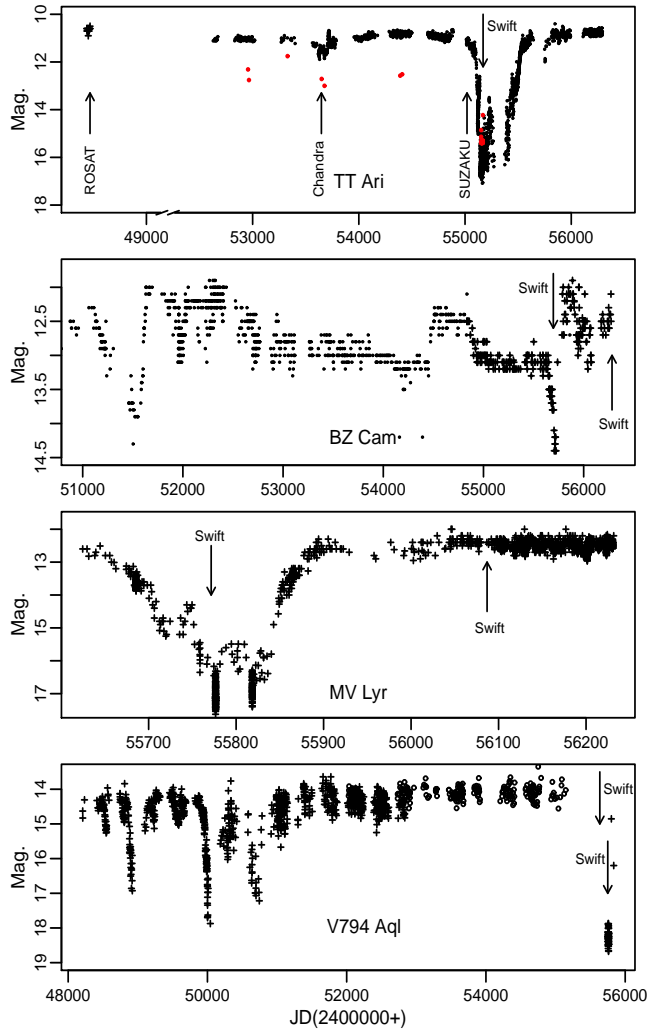}
\caption{Light curves of TT Ari, BZ Cam, MV Lyr and V794 Aql (from top to bottom)
obtained from the AAVSO (crosses), VSNET (filled circles) collaboration and ASAS 
(open circles) data. 
The times of the X-ray observations are marked with arrows.}
\label{fig:lc}
\end{figure}

\section{Previous optical and {\it UV} observations}
In Table 1 we report parameters from the published results of 
observations in the optical, near ({\it NUV}) and far ({\it FUV}) ultraviolet wavelength ranges.
We see that these objects have an orbital period just above the period gap, in a narrow 
range between 3.2 and 3.7 hours. For the three systems MV Lyr, TT Ari and V794 Aql the WD 
effective temperature T$_{\rm eff}$ was estimated in previous low states 
(not shown in Fig. 1) from {\it UV} and {\it FUV} observations, in the range 
 between 39,000 K and 47,000 K.
These  systems could not have been SSS at the time of those
observations, because T$_{\rm eff}$  places the flux peak in the {\it FUV} range. On the other 
hand, we cannot rule out ignition of thermonuclear burning, neither the possibility that 
the WD may become hotter with time in subsequent low states.
 The low state T$_{\rm eff}$ of MV Lyr tabulated in Tab. 1 was measured while the 
 {\it FUV} flux,
  corresponding to most of the bolometric one, was about 3.6 $\times 10^{-12}$ erg s$^{-1}$
 (see Table 1). In Fig. 5 of \citet{sta12} we see that at the tabulated distance these values
 may be consistent with thermonuclear burning with $\dot{m} < 1.6 \times 10^{-10}$ M$_\odot$
 yr$^{-1}$ and a WD mass less than 1 M$_\odot$. 
 \citet{godon11mvlyr} suggested that MV Lyr becomes hotter in the high state, 
 hypothesizing a lower limit T$_{\rm eff} \geq 50,000$ K, consistent with a measured 
 {\it FUV} 
 flux of 2.5 $\times 10^{-10}$ erg s$^{-1}$ in the high state. If the WD really becomes 
 hotter while emitting X-rays at this level, the possibility of thermonuclear burning
 would be even more likely in the high state.
 For a WD mass of 0.7 M$_\odot$ and the value of $\dot m$ in Table 1, assuming that 
 the extreme {\it UV} ({\it EUV}) luminosity is close to the total (bolometric) luminosity, nuclear 
 burning according to \citet{sta12} occurs with T$_{\rm eff}\simeq 80,000$ K
(see their Fig. 5), which implies a peak luminosity in the {\it EUV} range and no detectable SSS. 
 An accurate determination of T$_{\rm eff}$ in the high state is important: even
 an upper limit inferred from the absence of an SSS in the X-rays is useful to constrain 
 the evolutionary models.

\begin{table*}
\centering
\begin{minipage}{170mm}
  \caption{Binary parameters}
  \tabcolsep=0.1cm
  \begin{tabular}{l|cccc}
  \hline
														& BZ Cam 			& MV Lyr 										& TT Ari 								& V794 Aql \\
\hline
\hline
Dist(pc)												& $830\pm160^{[1]}$	& $505\pm50^{[3]}$ 								& $335\pm50^{[7]}$						& $690\pm105^{[10]}$\\
P$_{\rm{orb}}(d)$    									& $0.1536^{[2]}$	& $0.1329^{[4]}$    							& $0.1375^{[8]}$            			& $0.1533^{[11]}$\\
$i^{\rm{o}}$          									& 12--40$^{[1]}$ 	& 10--13$^{[4]}$     							& 17--22$^{[9]}$             			& $60^{[10]}$\\ 
M$_{WD}$(M$_{\odot}$) 									&               	& $0.7\pm0.1_{(FUV)}^{[3]}$   					& 0.57--1.2$_{(Opt)}^{[9]}$ 			& $0.88\pm0.39^{[10]}$\\
$\rm{\dot{m}}_{\rm{High}}$(M$_{\rm{\odot}}$ yr$^{-1}$)	&					&2--3$\times10^{-9}$$_{(FUV, Opt)}$ $^{[5],[6]}$& $1.1\times10^{-8}$$_{(Opt)}$$^{[9]}$ 	& 10$^{-8.5}$--10$^{-8.0}$$_{(FUV)}$$^{[10]}$ \\
$\rm{\dot{m}}_{\rm{Low}} $(M$_{\rm{\odot}}$ yr$^{-1}$)	&					&3$\times10^{-13}$$_{(Opt)}$$^{[3]}$			& 10$^{-16}$--$10^{-15}$$_{(UV)}$$^{[7]}$&  		\\
T$_{\rm{WD~High}}$    									&               	& $\geq50000 K_{(FUV)}^{[5]}$ 					&                         	 			&   	\\
T$_{\rm{WD~Low}}$     									&               	& $47000K_{(FUV)}^{[3]}$      					& $39000K_{(UV)}^{[7]}$    				& $45000K_{(FUV)}^{[10]}$ \\
T$_{\rm{Disk~Low}}$   									&               	& $<2500K_{(UV)}^{[6]}$       					&                          				&   	\\
{\it FUV} Flux$_{\rm{High}}$(erg cm$^{-2}$ s$^{-1}$)	& 					& 1.4$\times10^{-10}$$^{[5]}$$^*$				& $7.8\times10^{-10}$$^{[12]}$$^{**}$ 	& 		\\
{\it FUV} Flux$_{\rm{Low}}$ (erg cm$^{-2}$ s$^{-1}$)	& 					& 9.4$\times10^{-12}$$^{[3]}$$^{**}$			& 										& 		\\
\hline
\hline
\multicolumn{5}{p{.9 \textwidth}}{Notes:({\it FUV}), ({\it UV}), (Opt) -- values obtained from Far {\it UV}, {\it UV} and
Optical observations, respectively. $^*$ {\it FUV} flux was evaluated from the mean continuum level of a spectrum in a rage 910--1190 \AA.
$^{**}${\it FUV} flux was evaluated from the mean continuum level of a spectrum in a rage 920--1180 \AA.}\\
\hline
\multicolumn{5}{p{.9\textwidth}}{
[1]\cite{rin98bzcam}, [2]\cite{pat96bzcam}, [3]\cite{hoa04mvlyr}, [4]\cite{ski95mvlyr}, 
[5]\cite{godon11mvlyr}, [6]\cite{lin05mvlyr}, [7]\cite{gan99ttari}, [8]\citet{tho85ttari},
 [9]\citet{bel10ttari}, [10]\cite{god07v794aql}, [11]\cite{hon98v794aql}, 
 [12]\cite{hut07ttariFUV}}\\
\hline
\end{tabular}
\end{minipage}
\label{tab:par}
\end{table*} 

\section{Observations and data analysis}

We examined the archival X-ray data of VY Scl-type stars obtained with {\sl Swift},
{\sl ROSAT}, {\sl Suzaku} and {\sl Chandra} and chose the objects that were observed both
in the high and low states: BZ Cam, MV Lyr, TT Ari and V794 Aql.
The data are summarized in Table 2. All the data were never published except for a 
{\it ROSAT} observations of TT Ari, which we examined again and which were also analysed
by \citet{bay95ttariXrayQPO} and \cite{vantee96CVXray}. 

 In order to assess when the low and high optical states occurred, we relied on the data of the
 Variable Star Network (VSNET) Collaboration \citep{VSNET} \footnote{
 http://www.kusastro.kyoto-u.ac.jp/vsnet/}, the American Association
 of Variable Star Observers (AAVSO) \footnote{
http://www.aavso.org} and ASAS \footnote{
http://www.astrouw.edu.pl/asas/} databases. The optical light curves are presented
 in Figure \ref{fig:lc}. The epoch of the X-ray observation is marked with an arrow
 in each plot. We did not find optical data for V794 Aql around the epoch of the X-ray
 observations taken on 15 March 2011. However, from the photometric observations of
 the object before and after this date presented in \citet{hon14V794Aql}
it is reasonable to assume that V794 Aql was in the intermediate state
 during the X-ray observation.

 We used the  {\scriptsize HEASOFT} version {\scriptsize 6-13} to extract the {\sl Swift} and 
 {\sl Suzaku} spectra, and the  {\scriptsize XSPEC} version {\scriptsize 12.8.0} for spectral modelling.
  We also measured the {\it UV} magnitudes of the objects in both the {\sl Swift}/UVOT 
  observations and in additional {\sl GALEX} archival exposures. 
The {\sl Chandra} ACIS-S+HETG grating spectra were extracted with  {\scriptsize CIAO} 
version {\scriptsize 4.3}. Four different
 partial exposures were added for both the HEG and MEG gratings spectra with the 
  {\scriptsize FTOOL ADDASCASPEC},
 written originally for {\sl ASCA}, but also useful for all X-ray telescopes and detectors.

For better resolution and in order to use the full exposure, we also combined the data from
two {\sl Suzaku} front illuminated detectors (XIS 0 and XIS 3) taken in 3x3 and 5x5 modes. 
The timing analysis of the {\sl Suzaku} data was performed with the  {\scriptsize XRONOS}
 sub-package of  {\scriptsize FTOOLS} after the barycentric correction.

\section{Results}
\subsection{TT Ari}
TT Arietis is one of the brightest CVs, usually between {\it V} magnitude 10 and 11.
 Sometimes it abruptly falls into an `intermediate state' at {\it V}$\simeq$14 or even into a 
 `low state' reaching {\it V}$\simeq$18. According to 
\citet{bel10ttari} this binary system consists of a 0.57--1.2 M$_{\odot}$ white 
dwarf and a 0.18--0.38 M$_{\odot}$ secondary component of M 3.5$\pm0.5$ spectral type 
\citep{gan99ttari}.
The only low state before the one discussed in this paper was observed 
in the years 1980--1985 \citep{hud84ttari, sha85ttarilow}. 
The first panel of Figure \ref{fig:lc} shows the long term light-curve of TT Ari between
 1990 and 2013. The optical brightness started declining
 dramatically at the beginning of 2009 and the low state lasted for about 9 months,
 with a drop in optical luminosity of about 7 mag. However, in the low state the optical luminosity
 is not constant, with variations between {\it V}=15 and {\it V}=18. 

The high state X-ray spectrum of TT Ari was at first obtained by {\it EXOSAT} in Aug 21/22 1985 
\cite{hud87ttari}. Authors found that the X-ray flux in the range 0.2--4.0 keV was about
1.9$\times 10^{-11}$ erg cm$^{-2}$ s$^{-1}$. They also proposed that there are two or more 
hot emitting X-ray regions and two or more cold absorbing or scattering regions in TT Ari. 
On January 20 to 21, 1994 TT Ari was also
observed with {\sl ASCA} with an effective exposure time $\sim 14,000$ s. Detailed analysis
of these data was performed by \citet{bay96ttari}. One of the most interesting findings 
of the previous X-ray observations is
the rapid variability of the X-ray flux, a quasi-periodic oscillations (QPO) with a 
semi-period of 15--26 minutes (\citealt{bay95ttariXrayQPO}, \citealt{bay96ttari}). We will 
show that QPO with periods in this range are observed in all the high state observations
 we examined. In 2005 the {\sl Chandra} HETG spectra of TT Ari were obtained by C. Mauche 
(first shown in a presentation by Mauche, 2010).
Below we will discuss this set of observations in details.

\begin{table*}
 \begin{minipage}{130mm}
  \caption{Observational log of the {\sl ROSAT}, {\sl Swift}, {\sl Suzaku} and 
  {\sl Chandra} observations of VY Scl-type nova-likes that were analysed in this paper.}
\tabcolsep=0.1cm
  \begin{tabular}{clcccc}
  \hline
   Name & State & Date & Instrument & Exposure(s) & Count rate (cts s$^{-1}$) \\
\hline
\hline
BZ Cam 
& High & 21/12/2012   & {\it Swift}-XRT & 15001  & $0.0720\pm0.0024$\\  
& Low  & 15/05/2011   & {\it Swift}-XRT & 2710   & $0.100\pm 0.006$\\  
\hline
MV Lyr
& High & 08/06/2012  & {\it Swift}-XRT & 7569    & $0.096\pm 0.003$\\  
& Low  & 29/07/2011  & {\it Swift}-XRT & 3282    & $0.039\pm 0.004$\\ 
\hline
TT Ari
& High & 01/08/1991  & {\it {\sl ROSAT}}-PSPCB & 24464 & $0.098\pm 0.007$\\ 
& High & 06/09/2005  & {\it Chandra}-HEG & 95362$^*$ & $0.0266\pm 0.0003$\\   
& High & 06/09/2005  & {\it Chandra}-MEG & 95362$^*$ & $0.0588\pm 0.0005$\\
& High & 06/07/2009  & {\it Suzaku}-XIS FI$^{**}$  & 28617& $0.671\pm0.003$\\
& High & 06/07/2009  & {\it Suzaku}-XIS BI$^{**}$  & 28617& $0.836\pm0.005$\\
& Intermediate & 16/10/2009 & {\it Swift}-XRT    & 4421 & $0.278\pm 0.008$\\
& Low  & 22/11/2009  & {\it Swift}-XRT   & 12030 & $0.0285\pm 0.0019$\\
\hline
V794 Aql
& Intermediate & 15/03/2011  & {\it Swift}-XRT   & 6148  & $0.176\pm 0.005$\\ 
& Low  & 12/07/2011  & {\it Swift}-XRT   & 4629  & $0.055\pm 0.003$\\ 
\hline
\hline
\multicolumn{6}{p{.9 \textwidth}}{$^*$Four observations were taken with {\it Chandra}-MEG
and {\it Chandra}-HEG on September 6 and October 4, 6 and 9  2005.}\\
\multicolumn{6}{p{.9 \textwidth}}{$^{**}$ {\it Suzaku}-XIS FI -- are XIS 0 and XIS 3
detectors with front-illuminated (FI) CCDs, while {\it Suzaku}-XIS BI is the XIS 1 that 
utilizes a back-illuminated (BI) CCD}\\
\hline
\end{tabular}
\end{minipage}
\label{tab:obs}
\end{table*}

\subsubsection{The X-ray data: the high state}
A set of four {\sl Chandra} HETG exposures were obtained within 5 weeks in 
2005 (for details see Table 2). In the optical, the 
source was undergoing a `shallow decline'
 from the average optical magnitude in the high state, from {\it V}$\simeq$10.5
 to {\it V}$\leq$11.5. Fig. \ref{fig:ttarichandra} shows the coadded 
MEG and coadded HEG spectra of four subsequent
 eposures. There was no significant flux or spectral variability between
 the different exposures, and we shall describe
 the coadded MEG and HEG spectra. A rich emission line spectrum was measured, 
with strong emission lines of Mg, Si, Ne and S. 

\begin{figure}
\centering
\includegraphics[width=220pt]{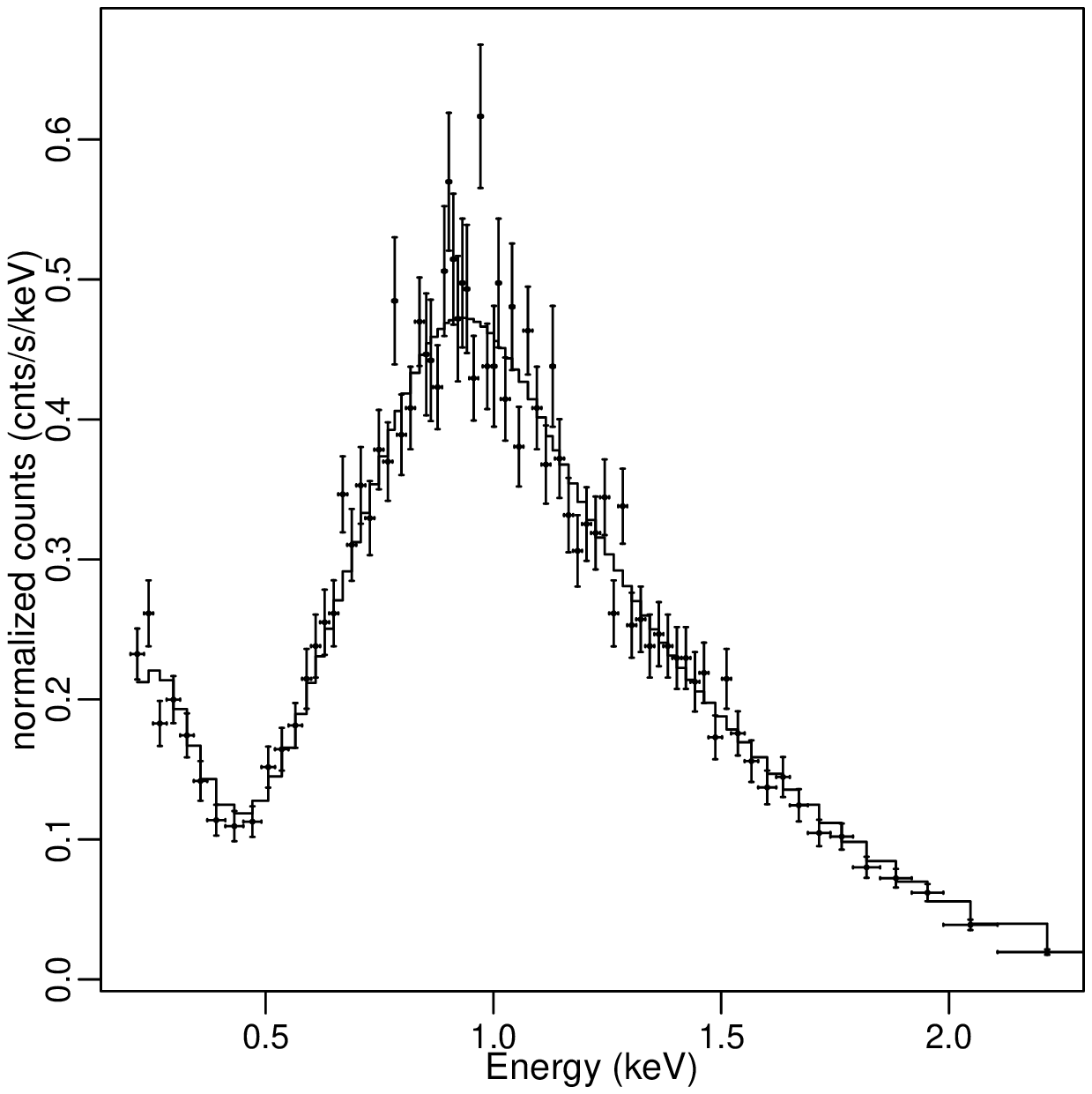}
\caption{High state X-ray spectrum of TT Ari taken with {\sl ROSAT}.}
\label{fig:ROSAT}
\end{figure}

 The strongest lines of the {\sl Chandra} MEG spectrum are listed in Table 3. For the 
 H-like lines we evaluated the flux with a Gaussian fit to the line; we also estimated the
  flux in the He-like triplet lines (Ne {\scriptsize IX}, Mg {\scriptsize XI}, 
  Si {\scriptsize XIII}), but we could only do it with larger uncertainty because the 
  lines are blended and we needed to fit three Gaussians (note that the intercombination 
  line is not resolved). Moreover, the triplets of He-like lines are observed 
 in a region of the spectrum which is rich in other lines, like those due to transitions
 of iron. Despite these difficulties, we performed the fit
 with three Gaussians for the triplets of Si {\scriptsize XIII} and Mg {\scriptsize XI}.
We added a fourth line of Fe {\scriptsize XVIII} at 13.509 \AA \ for Ne {\scriptsize IX}. 
 We thus evaluated the R ratio $f/i$ (intensity of the forbidden to the intercombination 
 line) and the G ratio $(f+i)\over r$ (where r is the intensity of the resonance line).    
 We estimated an uncertainty of about 20\% on both these ratios.
 We obtained R=0.63 and G=0.78 for Ne {\scriptsize IX}, R=0.36 and G=0.66 for Mg {\scriptsize IX}, R=0.33
 and G=0.66 for Si {\scriptsize XIII}.  
 
 \begin{figure*}
\centering
\includegraphics[width=160pt, height=510pt, angle=270]{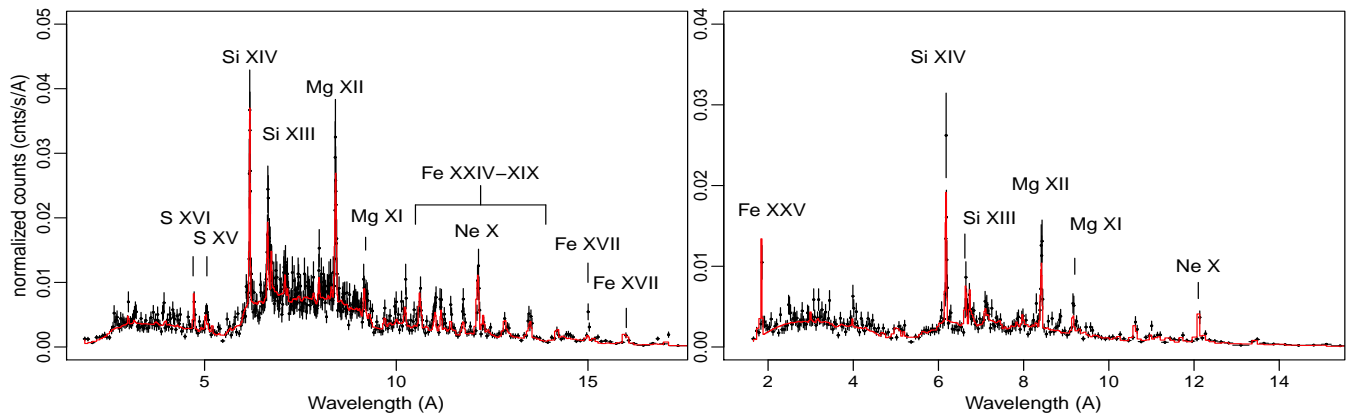}
\caption{TT Ari spectra observed with the {\sl Chandra} MEG (left) and HEG (right) grating. Four 
observations and +1 and -1 orders summed. The red lines represent 
the fit with two VAPEC components. The emission lines are indicated.}
\label{fig:ttarichandra}
\end{figure*}

 We consulted \citet{Porquet2000}, who explored the dependence of these ratios
 on electron density and plasma temperature. The authors assumed
 a photoionezed plasma, with or without
 additional collisional ionization. We see from their Figure 8 that the R ratios we obtained
 corresponds to high density; we obtain a lower limit on the electron density 
 n$_{\rm e}=10^{12}$ cm$^{-3}$,  $10^{13}$ cm$^{-3}$ and  $\geq 10^{14}$ cm$^{-3}$ for Ne, Mg and Si,
 respectively. However, it is known that the R ratios appears smaller,
 as if the density was higher than its actual value, when there is also
 photoexcitation by a strong {\it UV}/{\it EUV} source, exciting the $f$
 level electrons into the $i$ level \citep{Porquet2000}. We do expect additional photoexcitation 
  if the lines are produced very close to the 
 hot and luminous WD of TT Ari, thus, we do not know that
 in this case the R ratio is a completely reliable indicator. The G ratio, on the other 
 hand, is reliable and indicates plasma temperature $\geq 3 \times 10^6$ K,
 $\geq 4 \times 10^6$ K and $\geq 7 \times 10^6$ for Ne, Mg and Si, respectively. 

\begin{figure}
\includegraphics[width=230pt, height=260pt]{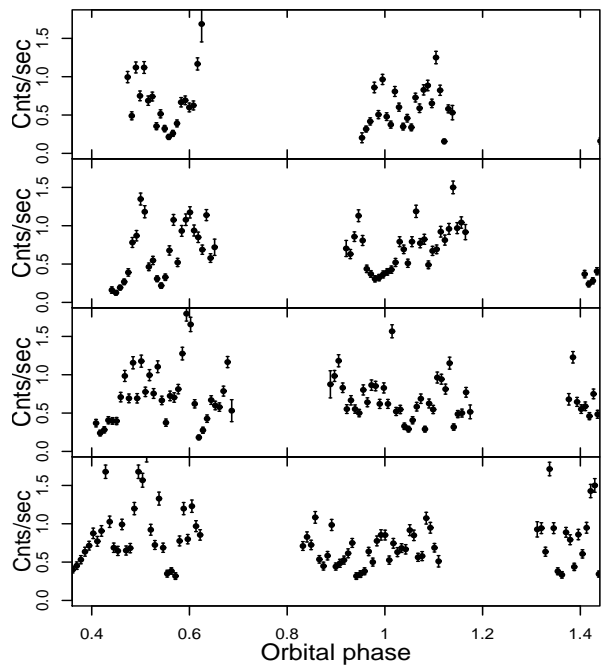}
\caption{X-ray light curve of TT Ari, obtained by {\sl Suzaku}, binned every 
100 seconds. The horizontal axis is the orbital phase. 
The plots are in chronological order.}
\label{fig:ttarixrlc}
\end{figure}

The next step was to fit the observed spectra with a physical model.
 A fit with a plasma in collisional ionization equilibrium is not acceptable because of too 
 high $\chi^2$, but
 adding a second temperature we obtained a more reasonable fit,
 with $\chi^2$=1.2. We adopted two BVAPEC models in  {\scriptsize XSPEC}, which describe the
emission spectrum of collisionally ionized diffuse gas, calculated using the ATOMDB code
 v2.0.1 with variable abundances at different temperature (see Table 4 
 and Fig. \ref{fig:ttarichandra}) and with velocity broadening. By letting the abundances
 of single elements vary, we found the best fit with the following values
for the abundances:
  [Ne/H]=$6.3\pm1.2$, [Mg/H]=$4.8\pm0.8$, [Si/H]=$4.9\pm0.9$, [S/H]=$11\pm4$, [Fe/H]=$1.46\pm0.25$,   
  [O/H]=$10\pm4$. The emission measure of the cooler component is 3.1$\times 10^{53}$ cm$^3$
 and the emission measure of the hotter component is 3.6 $\times 10^{54}$ cm$^3$.
 We note that, if these two regions are related to accretion,
 for an electron density of order of 10$^{14}$ cm$^{-3}$ (the minimum electron density
 derived from the G ratio for Si), the linear dimension of the emission region is of order of
 3.1 $\times 10^{8}$ cm and 7.1 $\times 10^{8}$ cm, respectively. This is, of course, an 
 purely phenomelogical model; two large and distinct regions with different
plasma temperature are difficult to explain in a physically realistic way. 
 
\begin{figure}
\centering
\includegraphics[width=240pt, height=270pt]{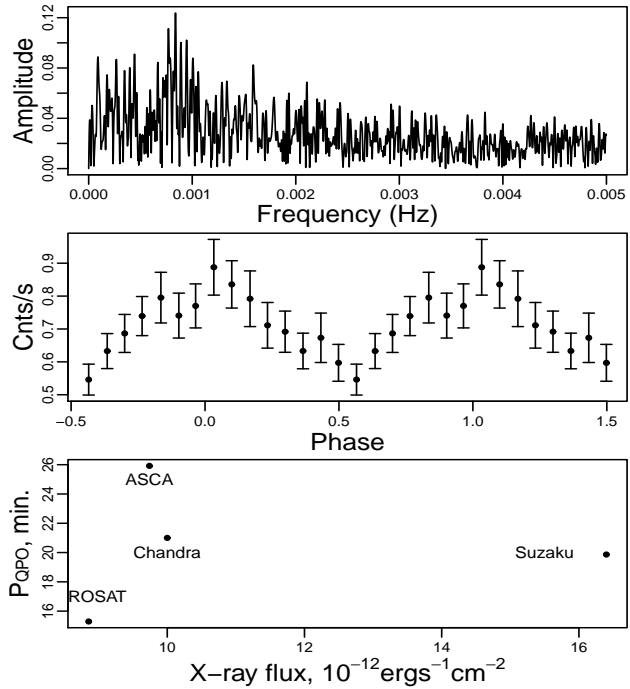}
\caption{Top panel: the Fourier power spectrum of the {\sl Suzaku} ligh curve of TT Ari.
The highest peak corresponds to 0.839 mHz or 19.8 min oscillations. Middle panel:
phase folded {\sl Suzaku} light curve with 19.8 min period. Bottom panel: periods of the
QPO and flux 
in the  X-ray range 0.5--10.0 keV in the {\sl ROSAT}, {\sl ASCA}, {\sl Chandra}
and {\sl Suzaku} observations. We assumed the values of \citet{bay96ttari} for the {\sl ASCA} 
observations and for the {\sl ROSAT} X-ray flux. The period of the QPO observed with 
{\sl ROSAT} was taken from \citet{bay95ttariXrayQPO}.}
\label{fig:fourier}
\end{figure}

BVAPEC model performs the Gaussian fitting of the lines and gives the $\sigma$ for two 
systems of lines associated with the two components (see Table 4). The full width at half 
maximum that corresponds to these values of $\sigma$ is about 1100 and 1500 km s$^{-1}$.

 We also wanted to try and better understand the physical scenario by adopting a more
 physically realistic model. \citet{muk03CVXray}
 have shown that accretion in all non magnetic CVs, and often even
 in magnetic ones, is best described by a stationary cooling flow model.
We thus used the cooling flow VMCFLOW model in 
{\scriptsize XSPEC} \citep[a cooling flow model after][]{mus88mkcflow}.
We see  however that this fit yields a larger $\chi^2$ value than the previous
 simplified model, and this is mainly because there is more flux in the He-like lines
 than predicted by the model. This may be due to additional
 photoionization, for instance in a wind from the system, 
 implying that not all the X-ray flux is produced in the accretion flow.
 The main problem, however, is that the cooling  
 flow model includes the mass accretion rate $\dot m$ as a parameter,
but predicts a very low value for $\dot m$, 
 only  $3.48 \times 10^{-11}$ M$_\odot$ yr$^{-1}$, while the
 UV and optical observations indicate  
 $10^{-8}$ M$_\odot$ yr$^{-1}$ for the high state $\dot m$ (see Table 1).
We thus conclude that either the observed X-ray flux does not originate in the accretion
 flow that produces the luminous accretion disk, or that accretion energy is mostly
 re-radiated in the {\it EUV} and not in the X-ray range.

\begin{table}
\begin{minipage}{90mm}
  \caption{Measured wavelength and fluxes in erg cm$^{-2}$ s$^{-1}$ $\times10^{-14}$ for the emission lines
  identified in the summed {\sl Chandra} MEG spectrum.}
  \begin{tabular}{|l|ccc|}
  \hline
Element  	& Energy$_{\rm{obs}}$ (KeV) & $\lambda_{\rm{obs}}$ ($\rm{\AA}$) & MEG flux$_{\rm{abs}}^{[1]}$\\
\hline
\hline	
S {\scriptsize XV}     	& 2.4606$^r$  					& 5.0387   						& 2.8\\
        				& 2.448$^i$  					& 5.064    						& 3.3\\
        				& 2.4260$^f$  					& 5.1106   						& 3.4\\
Si {\scriptsize XIV}   	& 2.0061$_{-0.0010}^{+0.0006}$	& 6.1803$_{-0.0017}^{+0.003}$  & 6.12\\
Si {\scriptsize XIII}  	& 1.8650$^r$  					& 6.6479   						& 3.4\\
        				& 1.854$^i$   					& 6.687    						& 1.7\\
           				& 1.8396$^f$  					& 6.6739   						& 0.57\\
Mg {\scriptsize XII}   	& 1.4733$_{-0.0011}^{+0.0003}$  & 8.4154$_{-0.0020}^{+0.006}$  & 3.07 \\
Mg {\scriptsize XI}    	& 1.3522$^r$  					& 9.1687   						& 2.0\\
            			& 1.3434$^i$  					& 9.2291   						& 1.6\\
            			& 1.3312$^f$ 				 	& 9.3136   						& 0.59\\
Ne {\scriptsize X}     	& 1.0211$_{-0.0004}^{+0.0007}$  & 12.142$_{-0.008}^{+0.005}$	& 5.25 \\
Ne {\scriptsize IX}    	& 0.9220$^r$  					& 13.44  	 					& 3.3\\
          		 		& 0.9149$^i$  					& 13.55  	 					& 1.8\\
           				& 0.9051$^f$  					& 13.69  	 					& 0.59\\
Fe {\scriptsize XVIII} 	& 0.8735      					& 14.19  	 					& 0.74 \\
Fe {\scriptsize XVII}  	& 0.8256$_{-0.0005}^{+0.0004}$  & 15.017$_{-0.008}^{+0.008}$  	& 2.98 \\
Fe {\scriptsize XVII}   & 0.7388$_{-0.0005}^{+0.0006}$  & 16.781$_{-0.013}^{+0.014}$  	& 4.8 \\
\hline
\hline
\multicolumn{4}{p{.9 \textwidth}}{[1] For a calculation of fluxes
in the lines we assumed N(H)=$0.06 \times10^{22}$. {\it r} -- resonance, {\it i} -- 
intercombination and {\it f} -- forbidden lines. }\\
\hline
\end{tabular}
\end{minipage}
\label{tab:lines}
\end{table}

 The light curves of these {\sl Chandra} exposures still show
 a QPO, although the modulation has a $\simeq$21 min period.
 We discuss in detail below the similar light curve we extracted from an 
 additional archival observation done with {\sl Suzaku}, which has higher S/N.

 {\sl Suzaku} observations of TT Ari were obtained by Saitou in 2009 just before the 
 low state (see top panel of Fig. \ref{fig:lc}). The average X-ray flux during this 
 observation was higher by almost a factor
 of 2 than during the {\sl Chandra} observation. In order to exclude the effect of a 
 slightly different energy ranges of the detectors we compared the X-ray flux in the 
 range 0.5--10.0 keV, common for both instruments.
The difference between the X-ray flux measured with {\sl Chandra} and {\sl Suzaku}
 may be correlated with the optical one. The {\sl Chandra} HETG
observations were taken at the time when TT Ari was optically less luminous ($\geq$ 1 mag).

\begin{figure}
\centering
\includegraphics[width=240pt, height=280pt]{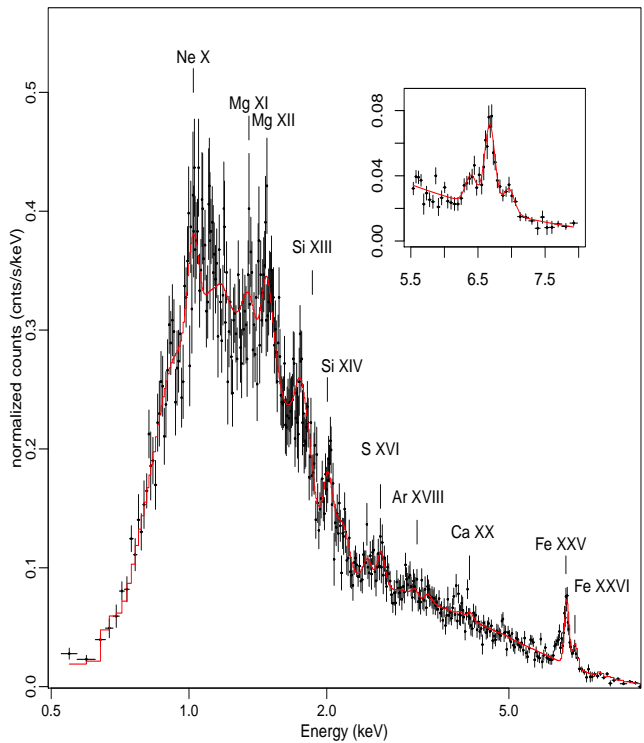}
\caption{Spectrum observed with the {\sl Suzaku} XIS FI (data from XIS0 and XIS3 detectors 
taken in 3x3 and 5x5 modes are summed). 
The red line shows the fit with two VAPEC components. The emission lines are indicated. The inset show 
the iron K$_{\alpha}$ emission lines at 6.41, 6.68 and 6.96 keV and their fit with three Gaussians.}
\label{fig:ttarisuzaku}
\end{figure}

The broad band spectrum of TT Ari observed with {\sl Suzaku} is presented in Fig. 
 \ref{fig:ttarisuzaku}. Emission lines of Ne, Mg, Si are clearly seen, like in the {\sl Chandra}
 spectrum, together with S, Fe {\scriptsize XXV}, and Fe {\scriptsize XXVI} lines. 
 We fit this spectrum either with a two-component thermal plasma model with temperatures 0.80 
 and 7.1 keV (for details see Table 3) and the abundances that were derived from the {\sl Chandra}
 spectra fit. The residuals of the fit indicate an extra line feature at 6.4 keV in 
 the {\sl Suzaku} spectrum that is most probably the iron fluorescence line. In the inset 
 in Fig. \ref{fig:ttarisuzaku} we show the fit of the 5.5--8.0 keV region with three
 Gaussians at 6.41, 6.68 and 6.96 keV. 
 The equivalent width of the K ${\alpha}$ iron line at 6.41 keV is $96_{-24}^{+24}$ eV. 
 This line implies that the X-ray emission region is close to a
 `cold' source, which may be the WD surface and/or an optically
thick accretion disk, and it is consistent with the WD being less hot than about 100,000 K. 

 The cooling flow model can be used also for the {\sl Suzaku} spectrum because we
 do measure spectral lines to constrain the model. The fit is not
 optimal, and we run into the same problem of low $\dot m$.

 The {\sl Suzaku} light curve is shown in Fig. \ref{fig:ttarixrlc} and is extremely 
 similar to the light curve previously observed with {\sl ROSAT} \citep{bay95ttariXrayQPO}. 
 The data were integrated in bins of 100 s. The QPO have an amplitude of 50 \%. 
 In Fig. \ref{fig:fourier} we present the Fourier 
 power spectrum of {\sl Suzaku} light curve of TT Ari. The highest peak corresponds to 
 0.839 mHz or 19.8 min oscillations. According to \citet{and09ttariatel2122}
 at the time of {\sl Suzaku} observations TT Ari also showed QPO in the optical range, with
 several peaks, at 2.5, 1.1 and 0.38 mHz. In \citet{bay95ttariXrayQPO} the frequency 
 of QPO in X-ray was explained as the beat frequency between the Kepler frequency at 
 the inner edge of the accretion disk and and the star's rotation rate. 
 Nevertheless, from the QPO semi-periods
 measured by us in the {\sl Chandra} and {\sl Suzaku} observations and from those 
 derived by \citet{bay95ttariXrayQPO} and \citet{bay96ttari}, 
 no correlation emerges between the observed frequency of QPO and the X-ray flux
of TT Ari (see the bottom panel of Fig. \ref{fig:fourier}).

In 1991 TT Ari was observed with {\sl ROSAT}. The {\sl ROSAT} X-ray spectrum is shown in 
Fig. \ref{fig:ROSAT}. \citet{bay95ttariXrayQPO} analysed these data and found that the
best fitting model is an absorbed black-body. We reanalysed the data and found that a 
black-body can represent only the soft part of the spectrum (the {\sl ROSAT} energy range 
is 0.2--2.4 keV) and for more sophisticated fit we need two-component thermal plasma model.
Parameters of the best fitting model are presented in Table 4. 

\subsubsection{The X-ray data: the low state}
In the intermediate and low state TT Ari was observed with {\sl Swift}. The 
observations in the intermediate state were presented by \cite{muk09ttariatel}. 
The unabsorbed flux was about the same as during the {\sl Suzaku} observation,
 1.5 $\times 10^{-11}$ ergs cm$^{-2}$ s$^{-1}$, the spectrum could only be
 fitted with a multi-temperature plasma, and a new quasiperiod of 0.4 days was also 
 measured in optical.
 
The low state spectrum, presented in the top panel of Fig. \ref{fig:highlowmod}, 
is best fitted with two components of absorbed
thermal plasma in collisional ionization equilibrium
 with a fixed metallicity APEC model 
at 0.7 keV and 3.9 keV respectively (we only set a fixed metallicity with 
 solar abundances because of the poorer data quality of this dataset). 
The low state X-ray flux appeared to be about ten times smaller
 than in the high and intermediate state,
 and definitely no luminous supersoft X-ray phase was detected.
 In Table 4 we show also present the comparison with only one component 
  at 3.4 keV. An upper limit to the temperature of a blackbody-like
 component is approximately 150,000 K. 

\begin{table*}
\centering
\begin{minipage}{170mm}
  \caption{Fitting models and parameters for TT Ari. The errors represent 90\% confidence region for a single parameter.}
  \tabcolsep=0.1cm
  \begin{tabular}{|lccccccc|}
  \hline
			& 	\multicolumn{5}{c}{high state}& \multicolumn{2}{c}{low state} \\
\hline 
\hline
Satellite       									& 	{\sl ROSAT} 					& 	\multicolumn{2}{c}{\sl Chandra} 							&	\multicolumn{2}{c}{\sl Suzaku}				& \multicolumn{2}{c}{{\sl Swift}}				\\
Models          									& 2 apec      				& 	2 bvapec				&	vmcflow		  			& 2 vapec+gauss$^*$	  		& vmcflow+gauss$^*$ 		&  2 apec 	 			& apec 					\\
\hline
N(H)$_1$($10^{22}$) 								&$0.0306_{-0.0026}^{+0.003}$&$0.048_{-0.048}^{+0.060}$	&$0.026_{-0.015}^{+0.013}$	& $0.136_{-0.021}^{+0.022}$ & $0.075_{-0.010}^{+0.007}$ & $0.04_{-0.04}^{+0.09}$& $0.019_{-0.019}^{+0.05}$\\
N(H)$_2$($10^{22}$) 								&			  				& $0.121_{-0.029}^{+0.029}$	& 			      			& 						 	&							&			 			&			   			\\
T$_1$ (keV)											& $0.89_{-0.11}^{+0.09}$	& $0.93_{-0.03}^{+0.029}$ 	& 				  			& $0.80_{-0.05}^{+0.13}$	& 							& $0.7_{-0.4}^{+0.3}$	& $3.4_{-0.7}^{+1.4}$	\\
T$_2$ (keV)											& $25_{-13}^{+25}$	  		& $6.5_{-0.4}^{+0.5}$   	& 			    			& $7.1_{-0.3}^{+0.3}$		& 				 			& $3.9_{-1.0}^{+2.7}$	& 						\\
Sigma$_1$ ($\rm{km s^{-1}}$)						&							& $646_{-81}^{+81}$			&							&							&							&						&						\\
Sigma$_2$ ($\rm{km s^{-1}}$)						&							& $461_{-105}^{+119}$		&							&							&							&						&						\\
EM$_1$ ($10^{53} \rm{cm^{-3}}$)$^{**}$				&							& $3.16_{-0.26}^{+0.43}$	&							&							&							&						&						\\
EM$_2$ ($10^{53} \rm{cm^{-3}}$)$^{**}$				&							& $36.50_{-0.10}^{+0.10}$	&							&							&							&						&						\\
T$_{\rm{min}}$ (keV)								&							&							& $0.20_{-0.026}^{+0.026}$	&							& $0.120_{-0.009}^{+0.016}$	&						&						\\
T$_{\rm{max}}$ (keV)								&							&							& $21.6_{-1.4}^{+1.9}$		&							& $26.9_{-1.5}^{+1.0}$		&						&						\\
$\rm{\dot{m}}$ ($\times10^{-11} \rm{M_{\odot} yr^{-1}}$)&						&							& $3.36_{-0.22}^{+0.27}$	&							& $5.02_{-0.11}^{+0.17}$	&			 			&						\\
Flux$_{\rm{abs}}^{***}$  							& $5.76_{-0.13}^{+0.3}$		& $9.35_{-0.27}^{+0.18}$	& $9.01_{-0.3}^{+0.011}$ 	&$15.81_{-0.020}^{+0.020}$	& $16.18_{-0.3}^{+0.20}$ 	& $0.99_{-0.18}^{+0.17}$& $0.94_{-0.18}^{+0.17}$\\
Flux$_{\rm{unabs}}^{***}$ 							& $6.70_{-0.13}^{+0.3}$ 	& $10.35_{-0.27}^{+0.18}$	& $9.21_{-0.3}^{+0.011}$ 	&$17.39_{-0.020}^{+0.020}$	& $17.19_{-0.3}^{+0.20}$ 	& $1.08_{-0.18}^{+0.17}$& $1.03_{-0.18}^{+0.17}$\\
$\chi^2$											& 	1.0	  		  			&	  1.2	    			& 		1.6 				&	    1.0					&	1.2						& 1.0		   			& 1.2  					\\
\hline  
\hline
\multicolumn{8}{p{.9 \textwidth}}{$^{*}$We added a Gaussian at 6.41 keV in order to fit the K${\alpha}$ iron reflection line in the {\sl Suzaku} spectrum.}\\
\multicolumn{8}{p{.9 \textwidth}}{$^{**}$Emission measure }\\
\multicolumn{8}{p{.9 \textwidth}}{$^{***}$The X-ray flux ($\times10^{-12}$erg cm$^{-2}$ s$^{-1}$) was calculated in the following energy ranges: 0.2--2.5 keV for {\sl ROSAT} PSPC, 0.4--10.0 keV for {\sl Chandra} HETG, 0.5--12.0 keV for {\sl Suzaku} XIS FI and 0.3--10.0 keV for {\sl Swift} XRT}\\
\hline
\end{tabular}
\end{minipage}
\label{tab:ttarimod}
\end{table*}

\subsubsection{UV data}
In the first panel of Figure \ref{fig:lc}, in addition to the optical
light curve of TT Ari, the red points show the {\sl GALEX}
 near {\it UV} ({\it NUV}) observations. In Table 6 
 we give exposure times and the mean AB magnitude in the {\it U}/{\it UV} filters during the low 
 and high states. 
The amplitude of the low to high state transition in {\it NUV} was much lower than in optical: 
 3 versus 7 magnitudes. Like in the optical range,
TT Ari shows flaring activity in the {\it NUV}, 
with amplitudes up to 1 mag. However, the {\it UV} and optical flares occur
 at different times, and do not appear to correlated, neither anti-correlated. It may
be due to the different origine of the {\it UV} and optical radiation. 

\subsection{BZ Cam}
BZ Cam shows brightness variations around an average value {\it V} = 12--13, with rare 
occasional transitions to low states with {\it V} = 14--14.5. Besides the low state studied
 here, two additional low states were detected -- in 1928 and at the beginning of 2000 
(\citealt{gar88vyscldqher} and \citealt{gre01bzcam}, respectively). 
BZ Cam is surrounded by a bright emission nebula with a bow-shock structure, 
 first detected by \cite{ell84PNsurvey} and also studied by \cite{kra87bzcam}, 
\cite{hol92bzcambowshock}, \cite{gre01bzcam}.
\cite{hol92bzcambowshock} proposed that the bow shock structure is 
 due to the interactions of the wind of BZ Cam with the interstellar medium. 
The wind in BZ Cam was also studied by \cite{hon13bzcam}.
\citet{gre01bzcam} suggested that this nebula is photoionized by a bright 
 central object that must be a super soft X-ray source, while the
 bow shock structure is due to the high proper motion of BZ Cam, moving
 while it emits the wind. 
 
\subsubsection{X-ray data}
The second plot of Fig. \ref{fig:highlowmod} shows the X-ray spectra of BZ Cam observed 
with the {\sl Swift} XRT. 
The luminosity is higher in the low state, however, in the very soft spectral region,
at energy $\leq$0.5 keV, the X-ray flux is almost twice higher in the high state,
 which is exactly the opposite of the scenario predicted 
 by \citet{gre99v751cyg}. Interestingly, the spectral fits in both states 
indicate that we may be observing an unresolved, strong Ne {\scriptsize X} Lyman ${\alpha}$
 line at 1.02 keV. In the high state spectrum the Fe {\scriptsize XXV} line at 6.7 keV is 
 clearly detected. BZ Cam was also previously observed with {\sl ROSAT} in the high state. 
\cite{vantee94CVROSAT} and \cite{gre98vyscl} fitted the spectrum 
either with one component blackbody or with a highly absorbed bremsstrahlung (or MEWE) 
model. \cite{gre98vyscl} favored the blackbody. However, with the larger energy range 
of {\sl Swift} we see that a fit including also the high energy part of the spectrum, 
 is only possible with at least two components, and that the blackbody is not adequate.
The black dots in the second panel of Fig. \ref{fig:highlowmod} show the low state 
spectrum of BZ Cam, and a fit with a two-component VAPEC model.
 With only a broad band spectrum and no detected emission lines, we could not adequately 
 fit the cooling flow model.
 The same is true for other low resolution X-ray spectra described below.

The high state of BZ Cam (red dots in the second panel of Fig. \ref{fig:highlowmod}) is 
best fitted with a two components VAPEC model. The fitting parameters are listed in Table 
5. We find the best fit with non solar abundances,
[Fe/H]=($4.5\pm2.4$) and [Ne/H]=($\sim14$). In the low state, the increased flux
 seems to be due to a higher maximum temperature, 64 keV instead
 of about 10 keV. The column density N(H) appears to increase in the low state.
 Again, a lower limit to the  blackbody temperature is of about 150,000 K.

\begin{table*}
\begin{minipage}{120mm}
  \caption{Fitting models and parameters for BZ Cam and MV Lyr. The errors represent 90\% confidence region for a single parameter.}
  \begin{tabular}{l|ccc|ccc}
  \hline
 						& \multicolumn{3}{c}{BZ Cam}	   											& 						\multicolumn{3}{c}{MV Lyr}			\\
 						& \multicolumn{2}{c}{high state}   			  		& low state 			& \multicolumn{2}{c}{high state}  				& low state  \\
\hline
\hline
Satellite       		& \multicolumn{2}{c}{\sl Swift}    			  		& {\sl Swift}			& \multicolumn{2}{c}{\sl Swift}   				& {\sl Swift}\\
Models          		&  vapec		  			&  2 vapec   			& 2 apec  				& vapec+plow 			& 2 apec     			& 2 apec \\
N(H)($10^{22}$) 		&  $0.14_{-0.03}^{+0.03}$ 	& $0.26_{-0.08}^{+0.12}$& $0.7_{-0.7}^{+0.4}$	& $0.24_{-0.24}^{+0.5}$	& $0.69_{-0.26}^{+0.16}$& $0.73_{-0.52}^{+0.73}$ \\
Photon Index    		&				  			&				 		&		    			& $1.00_{-0.24}^{+0.5}$	& 	 			  		&     					\\
T$_1$ (keV)     		&  $9.3_{-1.7}^{+0.2.8}$  	& $0.51_{-0.20}^{+0.3}$ &$0.3_{-0.3}^{+0.0.4}$  & $0.18_{-0.49}^{+0.38}$& $0.11_{-0.04}^{+0.03}$& $0.12_{-0.05}^{+0.07}$ \\
T$_2$ (keV)     		&  	              			& $10.8_{-2.3}^{+4}$  	&$63.6_{-54}^{+0.4}$ 	&     					& $58_{-50}^{+6}$ 		& $2.47_{-1.6}^{+1.9}$ \\ 
Flux$_{\rm{abs}}$$^*$   &$3.85_{-0.29}^{+0.24}$		&$4.04_{-0.3}^{+0.22}$ 	&$6.4_{-2.2}^{+5}$		& $6.5_{-1.6}^{+0.4}$ 	& $6.0_{-6}^{+3.5}$ 	& $1.2_{-0.4}^{+1.3}$ 		\\
Flux$_{\rm{unabs}}$$^*$ &$4.24_{-0.29}^{+0.24}$		&$4.98_{-0.3}^{+0.22}$  &$14_{-2.2}^{+5}$		& $9.7_{-1.6}^{+0.4}$ 	& $90_{-90}^{+52}$     	& $75_{-25}^{+81}$ 			\\
$\chi^2$        		&  1.6  		  			& 1.1			  		& 1.0					& 1.0 					&  1.2 		 			&    						\\
\hline
\hline
\multicolumn{7}{p{.9 \textwidth}}{$^*$The X-ray flux ($\times10^{-12}$erg cm$^{-2}$ s$^{-1}$) for {\sl Swift} XRT was calculated in the range 0.3--10.0 keV}\\
\hline
\end{tabular}
\end{minipage}
\label{tab:bzmvmod}
\end{table*}

\subsubsection{UV observations}
\begin{figure}
\includegraphics[width=220pt]{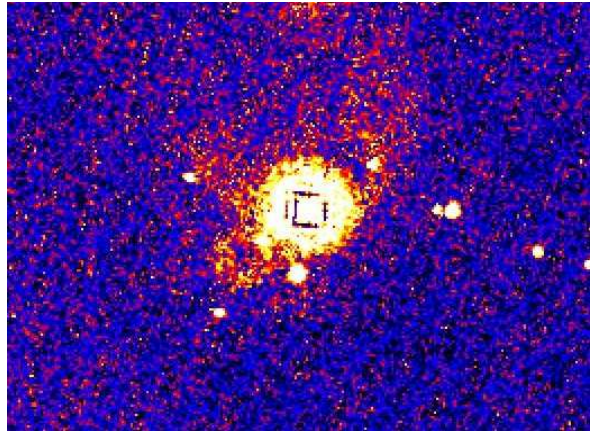}
\caption{{\sl Swift} UV image of BZ Cam, notice the emission nebula.}
\label{fig:bzcamim}
\end{figure}

BZ Cam was observed in {\it UV} with {\sl Swift} during the low state with a 1172.43 s exposure
and with a 121 s exposure during the high state.
The UV magnitudes in the AB system in the high and 
low states, given in Table 6, indicate a smaller variation 
than observed in the other objects.
 This is explained by Figure \ref{fig:bzcamim} in which we show the UV image of
 the nebula obtained with {\sl Swift}/UVOT observations. Even with the poor
 spatial resolution of UVOT, we detect an extended object; 
obviously the ionized nebula also emits copious UV flux.
 Comparing the image of BZ Cam in Figure \ref{fig:bzcamim} and the optical image 
 in Figure 4 of \cite{gre01bzcam}, we see a trace of
 the bow shock oriented in the South-slightly South-West direction,  but there seems to be 
additional {\it UV} emission of the nebula in the North-East direction,
 unlike in the in O {\scriptsize [III]} and H ${\alpha}$ images taken in September of 2000
 with the WIYN telescope \citep{gre01bzcam}.
 
\begin{table*}
\begin{minipage}{120mm}
  \caption{UV observations}
  \tabcolsep=0.05cm
  \begin{tabular}{lcccccc}
  \hline
  Object & State & Date & Instrument & Exp.(s) & Mag.$_{AB}$ & Filter$^*$\\
  \hline
  \hline
  BZ Cam
  & High 		& 	21/12/2012			& {\sl Swift} UVOT & 1200.74    & $13.0\pm 1.0$     & UVW2 \\
  & Low  		& 	15/05/2011			& {\sl Swift} UVOT & 2737.32    & $13.198\pm 0.011$ & U    \\
  \hline
  MV Lyr 
  & Low  		&	29/07/2011			& {\sl Swift} UVOT & 2092.81   & $16.265\pm 0.019$ & U    \\
  \hline
  TT Ari
  & High 		& 	01/11/2005			& {{\it GALEX}}     & 1468.3	 & $12.52\pm 0.001$ & FUV  \\
  & High 		&13/11/2003--03/11/2007 & {{\it GALEX}}     & 80--1686   & $12.518\pm 0.003$ & NUV  \\
  & Low  		&15/11/2009--02/12/2009 & {{\it GALEX}}     & 1466--1702 & $15.364\pm 0.003$ & NUV  \\
  & Low  		& 22/11/2009			& {\sl Swift} UVOT  & 6311.28    & $15.091\pm 0.011$ & UVW2 \\
  \hline
  V794 Aql
  & Intermediate& 	15/03/2011			& {\sl Swift} UVOT & 6186.67    & $17.073\pm 0.025$ & UVW1 \\
  & low  		& 12/07/2011			& {\sl Swift} UVOT & 4732.45    & $19.26 \pm 0.10$  & UVM2 \\
  \hline
  \hline
\multicolumn{7}{p{.9\textwidth}}{{\sl Swift} filters central wavelengths (\AA): U -- 3465, UVW1 -- 
2600, UVM2 -- 2246, UVW2 -- 1928. }\\
\multicolumn{7}{p{.8\textwidth}}{ {{\it GALEX}} UV band (\AA): NUV -- 1750--2800, FUV -- 1350--1750 } \\
\hline
\end{tabular}
\end{minipage}
\label{tab:uv}
\end{table*}

\subsection{MV Lyr}
MV Lyr spends most of the time in the high state, having brightness 
in the range {\it V} = 12--13, and during the occasional short low states it is in the range 
{\it V} = 16--18. Historical light curves of MV Lyr can be found in \cite{ros93mvlyr}, 
\cite{wen93mvlyr}, \cite{and82mvlyr} and \cite{pav98mvlyr2}.
With their {\sl FUSE} (Far Ultraviolet Spectroscopic Explorer) observations 
\cite{hoa04mvlyr} estimated that
 during the low state $\dot m \leq 3\times10^{-13}$ M$_{\odot}$ yr$^{-1}$, a four orders 
 of magnitude decrease from the value of $\dot m$ estimated by \cite{godon11mvlyr} in the 
 high state.

\subsubsection{X-ray data}

\begin{figure}
\centering
\includegraphics[width=230pt, height=390pt]{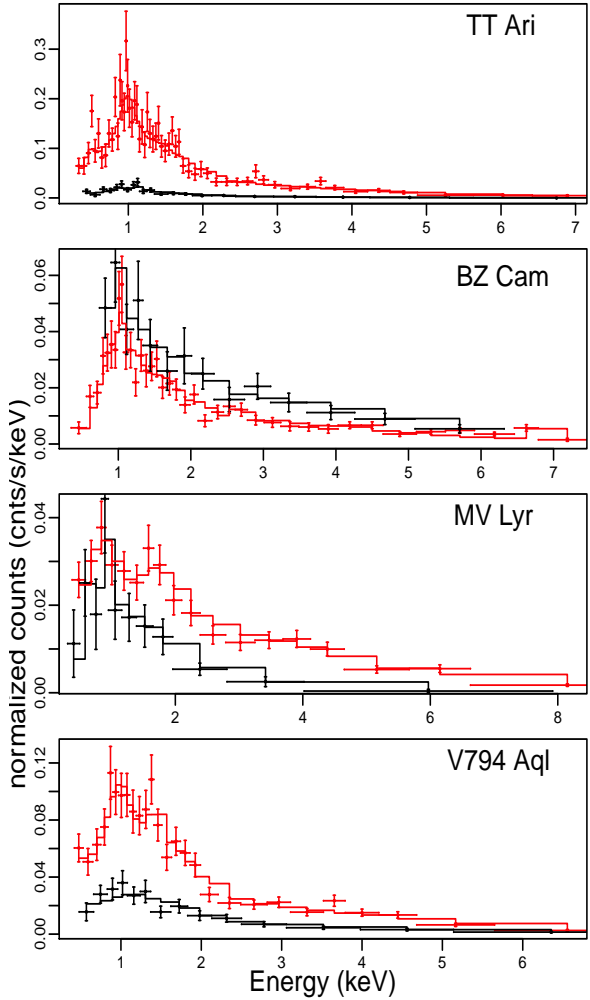}
\caption{The low and high state X-ray spectra of BZ Cam and MV Lyr 
observed with {\sl Swift}. TT Ari and V794 Aql were observed with {\sl Swift} during the low
and intermediate state. 
The high and intermediate state spectra are plotted
 in red and the low state spectra in black.
The solid lines show the models, the dots with error bars
 indicate the data.}
\label{fig:highlowmod}
\end{figure}

The third plot of Figure \ref{fig:highlowmod} shows the comparison between the high and 
low state X-ray spectra of MV Lyr obtained with {\sl Swift }together with the spectral 
fits. Unlike in BZ Cam, the high 
state X-ray flux of MV Lyr is higher by an order of magnitude than in the low state. The 
spectrum is also harder, with an additional component prominent above 1.7 keV.

We fitted the high state spectrum of MV Lyr with a two components thermal plasma model,
 but a good fit is also obtained with a thermal plasma and a power law model. 
 We fitted the low S/N, low state spectrum with a two components plasma model with
 fixed solar abundances, without any attempt to explore the role of the abundances (see Table 5).

A {\sl ROSAT} observation of MV Lyr in November 1992 in the high state was studied by 
\cite{gre98vyscl}. The authors fitted the spectrum with a black body,
 however, like in the case of BZ Cam, this model fails to fit the high energy part of the 
 spectrum that we measured with {\sl Swift}.
\cite{gre98vyscl} observed MV Lyr at the end of the 9 week optical 
low state in 1996 and obtained only an upper limit
 for the X-ray luminosity of 10$^{29.7}$ erg s$^{-1}$ assuming
 a distance of 320 pc (smaller than
 the most current estimate of 505$\pm$50 pc
 we give in Table 1). Assuming a distance of 320 pc, the flux
 measured during low state observation done with {\sl Swift} (see Tab. 5)
four months after the beginning of the decline to the low state,
one month after minimum, would be 10$^{31}$ erg s$^{-1}$, more than
 a factor of 10 higher than this upper limit. 
Thus, it seems that the X-ray flux of MV Lyr in the low state is not quite constant.

\subsection{V794 Aql}
In the high optical state V794 Aql varies between 14th and 15th magnitude, and in the low
 states it can plunge to 
18--20 mag. \citep[in the {\it B} filter, see][]{hon85v794aql}.
 \cite{god07v794aql} fitted spectra of the Hubble Space 
Telescope Space Telescope Imaging Spectrograph ({\sl HST-STIS}) and of {\sl FUSE}. They 
derived the following binary system parameters: M$_{\rm{WD}}=0.9$ M$_{\odot}$, 
 high state $\dot{\rm{M}}$ = $10^{-8.5}-10^{-8.0}$ M$_{\odot}$ yr$^{-1}$, inclination $i = 60^o$, 
 and distance to the system d = 690 pc. 
 
\subsubsection{The X-ray data}
The spectra and their fits of V794 Aql in the intermediate 
state ({\it V}$\simeq$15.5) and in the low 
state are presented in the bottom panel of Fig. \ref{fig:highlowmod}. The X-ray flux 
is three times higher in the intermediate than in the low state. We fitted the intermediate
 state spectrum of V794 Aql with two VAPEC components 
(see Table 7). In both components we need high abundance of Mg ($\sim5$). Again, 
there is no luminous supersoft X-ray source in the low state,
 with an upper limit of about 150,000 K.

\begin{table}
\centering
\begin{minipage}{80mm}
  \caption{Fitting models and parameters for V794 Aql. The errors represent 90\% confidence region for a single parameter.}
  \begin{tabular}{|l|cc|}
  \hline
						& high state   						 & low state \\
\hline 
\hline
Satellite       		& {\sl Swift}  						 & {\sl Swift} \\
Models          		& 2 vapec	  						 & apec \\
N(H)($10^{22}$) 		& $0.05_{-0.03}^{+0.04}$			 & $0.005_{-0.05}^{+0.07}$\\
T$_1$ (keV)				& $0.9_{-0.3}^{+0.3}$	 			 & $8_{-3}^{+20}$ \\
T$_2$ (keV)				& $16_{-8}^{+39}$					 &  \\
Flux$_{\rm{abs}}$$^*$	& $8.2_{-1.3}^{+0.5} \times10^{-12}$ & $2.5_{-0.6}^{+0.4} \times10^{-12}$ \\
Flux$_{\rm{unabs}}$$^*$	& $8.8_{-1.3}^{+0.5} \times10^{-12}$ & $2.7_{-0.6}^{+0.4} \times10^{-12}$ \\
$\chi^2$				& 1.0	 	  			 			 & 0.7 \\
\hline  
\hline
\multicolumn{3}{p{.9 \textwidth}}{$^*$The X-ray flux ($\times10^{-12}$erg cm$^{-2}$ s$^{-1}$) for {\sl Swift} XRT was calculated in the range 0.3--10.0 keV}\\
\hline
\end{tabular}
\end{minipage}
\label{tab:v794mod}
\end{table}

\section{Discussion}

An important motivation for this research has been the claim by \cite{gre98vyscl} and 
\cite{gre01bzcam} that some of the WD in VY Scl-type stars must be burning hydrogen 
quietly in the low state, without ever triggering thermonuclear flashes because of the 
short duration of the burning. We found that the predicted supersoft X-ray source does not
 appear in the low states, so thermonuclear burning at high temperature is ruled out.
 However, as we see in Table 1 at least 3 of the 4 objects we investigated have low mass 
 M$_{\rm{WD}}$. According to \citet{sta12} thermonuclear burning in WDs 
 whose mass is lower than 1 M$_\odot$ occurs quietly with atmospheric temperature below 
 150,000 K, outside of the SSS window, except for very high values of the mass accretion 
 rate (see Fig.5 of these authors).
 For Wolf et al. we see that WDs with mass up to 0.7 M$_\odot$
 burn hydrogen in the stable regime (without nova
 eruptions) with $\dot{m}$ of a few times
 10$^{-7}$ M$_\odot$ year$^{-1}$ have T$_{\rm eff}\leq$200,000 K.
 We note that also CAL 83, hosting a WD burning H at a much higher atmospheric 
 temperature, has low, intermediate and high states in the optical and X-rays, although 
 with a smaller amplitude in the optical than the VY Scl binaries. These variations were
 associated with the changes in the amount of irradiation of the accretion disc
 (see e.g. \citealt{gre02CAL83}, \citealt{raj13CAL83}). 
 
All the X-ray spectra of the VY Scl systems we examined appear complex, and the 
{\sl Chandra} and {\sl Suzaku} spectra of TT Ari clearly indicate more than one emission 
region or mechanism. The best-fitting model for all the 0.3--10.0 keV broad band spectra 
is a, probably still simplistic, two-component absorbed thermal plasma model.
Here we discuss the possible origins of the observed X-ray emission. 

\subsection{Accretion disk boundary layer}
We found that in three of the systems the X-ray luminosity decreases during the optical and 
UV low states, although the X-ray flux variation is the smallest.
The X-ray flux seemed to be anti-correlatd with the optical and UV only in 
 BZ Cam. Thus, if in the other three systems the main
 source of {\it UV}/{\it FUV} and optical luminosity is the accretion disk, 
 it seems unlikely that the X-ray flux is due to the innermost
 portion of the disk. In fact, our fits with the 
 cooling flow model, which generally yields good results assuming that all the X-ray flux 
 is emitted in an accretion flow, for all four
 objects return unreasonably low values of $\dot{m}$, which cannot be reconciled with 
 measurements at other wavelengths. This is not completely unexpected, since the accretion
 disks of systems accreting at high $\dot m$, close to 10$^{-8}$ M$_\odot$ yr$^{-1}$, 
 seem to re-radiate mostly or completely in the {\it EUV} range \citep{Pop95BL}, 
 because the boundary layer is optically thick.

For TT Ari, the semi-regular variability (QPO) with periods of 17--26 minutes
in the high state is best explained with the flickering of an accretion disk, 
 however we also found that there is no correlation between the X-ray flux and the 
 frequency of the QPO, which would be expected for accretion disk flickering 
 (\citealt{bay95ttariXrayQPO}, \citealt{pop99BL}).
 
\subsection{X-ray emission in a  wind}
 If the origin of the X-ray emission is not in the boundary layer of the accretion 
 disk, it may originate in a wind, either from the WD or from the accretion disk, depleting
  matter from the system. Such a wind may play an important role in the evolution,
  preventing the WD from reaching the Chandrasekhar mass.
  The fit of the TT Ari emission lines observed with {\sl Chandra} indicates a FWHM in the
   range 1100--1500 km s$^{-1}$. 
 However, the lines do not display any measurable blue or red shift to prove a wind scenario. There is
 significant broadening, but it may be due to collisional 
 ionization in the accretion flow, or to matter in almost-Keplerian rotation.
The WD effective temperature and {\it FUV} flux reported in Table 1 are consistent with a
line driven wind, although if nuclear burning takes place, the radius of WD 
  at some stage may increase, and we cannot rule out that at some (still not observed) 
  brief stage the WD reaches a luminosity where also electron 
  scattering opacity starts playing a role (a nova-like radiation pressure wind). 
  In either case, the most likely origin for the X-ray flux in the
 observations we examined is circumstellar material, shocked when it collides
 with a new outflow, possibly at a large distance from the WD. 
 There may be circumstellar material left from the AGB phase of primary or old 
  remnant of a previous nova, or a previous `thicker' wind caused by enhanced luminosity 
  due to nuclear burning, that has slowed down. A strong stellar wind is very likely to 
  play a role in the extended BZ Cam nebula, which was initially classified as a planetary nebula.
Instead we would argue that for TT Ari this explanation cannot account for the largest portion
 of the X-ray flux, because this system shows a 6.4 keV reflection line, which indicates that
a large fraction of X-rays (at least X-rays above 7 keV) must originate close to the
white dwarf or to the disk. 

There is a secure observation of X-rays far away from the accretion disk in UX UMa 
(see \citealt{pra04UXUMa}), an eclipsing nova-like with a hard, absorbed, eclipsed 
X-ray component and a soft, unabsorbed, uneclipsed X-ray component. 
The soft X-rays in UX UMa may indeed originate in a wind from the system.
 A fast wind is also known to occur in CAL 87 (\citealt{gre04CAL87}, \citealt{ori04CAL87}),
another system that may be closely related to the VY Scl-type stars. 
 The X-rays and optical flux variations anticorrelate only in BZ Cam, so it is possible that
 in this system the wind increases in the low state, causing additional absorption
 and obscuring the accretion disk. 
 
 Disk winds are observed in many types of 
compact objects, while the mechanism that causes them is not completely clear.  
At optical and UV wavelengths a mass outflow from the disk has been inferred 
in some CV via
 the observation P Cygni profiles, most notably the C{\scriptsize IV} $\lambda$1549 \AA 
~line (\citealt{rob73ZCam}, \citealt{cor82ZCam}, \citealt{lon02wind}).
P Cygni line profiles or/and absorption features have also been detected in X-rays in low mass 
X-ray binaries (\citealt{ued01FeK}, \citealt{bra00CirX1}) and are assumed to originate in a
high-velocity outflow from a flared and X-ray-heated accretion disk.
Disk winds also cause additional circumstellar, sometimes time-dependent, absorption 
components in the soft X-rays in non-magnetic CV's (\citealt{bas01wind}, \citealt{sai12wind}).

\subsection{Polar caps}
 A tempting hypothesis is that, while one component of the X-ray flux is due to a mass 
 outflow from the system, another component originates in a different, and coexistent mode
  of accretion other than the disk, i.e. a stream to the polar caps. In short, the VY Scl 
  would be intermediate polars (IP's). This scenario
 explains the lack of a clear correlation of {\it UV}/optical versus X-ray flux variations.
  As in the model proposed by \citet{ham02vysclmagnetic} the stream to the polar caps still
   continue, at decreased rate, when the accretion disk is periodically disrupted in the 
optically low state. In an IP, the disk would emit in optical and {\it UV}, but it
 would be truncated instead of having an X-ray emitting boundary layer, no matter what the
 value of $\dot m$ is.

Mauche (2010) compared spectra of magnetic and non-magnetic CV's and made a point that
 division into the two classes is not clear-cut on the basis of the X-ray spectrum alone,
 because of the large variety of observed X-ray spectra of magnetic CV's. There is no
 `typical' spectrum among polars and IP's. There is evidence for and against
 the magnetic scenario for TT Ari, but the X-ray spectrum alone does not prove or disprove
 it.

 An X-ray flux modulation due to the WD rotation period, which is 
 very typical and is considerd the smoking gun to classify IP's, has not been detected in 
 these systems so far. For three of them the reason may be low inclination, but not so for 
 V794 Aql. However, if the major component of the X-ray flux in the high state is not the 
 accretion stream to the poles, but it is associated instead with a wind, 
 isolating the 
 accretion component for the timing analysis is a serious hurdle in detecting a 
 periodicity due to the WD rotation.

\section{Conclusions} 
The VY Scl binaries are critical to understand the evolution
 of WD interacting binaries. They pose several riddles for the theories and understanding
 them well is a key to a consistent evolutionary picture. 
 Are these systems almost always quietly accreting at a high rate, with short intervals of low
 $\dot m$ that prevent the occurence of a thermonuclar flash and 
 mass loss in nova outbursts? Is thermonuclear burning of hydrogen on-going
at all phases, and how do we find evidence since we do not observe their WD at the high 
effective temperature necessary to emit supersoft X-rays?

We analyzed a number of X-rays and UV observations of four VY Scl systems comparing
 phenomena occurring during the optically `high' and `low' state. We did not detect 
 supersoft X-ray emission in both states, however, we cannot exclude 
 H burning at a lower temperature, outside of the SSS window, as can be expected 
 from the low masses of the white dwarfs in these systems. The data collected and 
 examined in this paper suggest that the X-ray emission has more than one component
 in all the four systems.
 We concluded that one component most likely originates in the circumstellar material, 
 shocked by the wind, possibly at a large distance from the WD while the second component 
 can be X-ray emission from the polar caps. However, we are not able to prove neither
 clearly disprove an IP scenario for these systems. 
 
  It can be argued that the X-ray observations
 at this stage have posed more new puzzles. We suggest that monitoring these systems over 
 the years in optical, {\it UV} and X-rays as frequently and simultaneously as possible is   
 a key to understand how accretion occurs and how it interplays with the thermal state
 of the secondary.  More intensive monitoring, that may be done with {\sl Swift}, 
 would be very rewarding, allowing to understand whether an evolutionary path
 at high mass transfer rate without mass loss in nova outbursts 
 can be sustained for a long time, and whether it leads to `quieter' outflows
 preventing the WD growth in mass, or to evolution towards a type Ia supernova. 
	
\section*{Acknowledgments}
Polina Zemko acknowledges the grant of the National Scholarship Programme SAIA and a 
pre-doctoral grant of the CARIPARO foundation at the University of Padova.
Dr. Shugarov acknowledges the VEGA grant No.2/0002/13.

\newcommand*\aap{A\&A}
\let\astap=\aap
\newcommand*\aapr{A\&A~Rev.}
\newcommand*\aaps{A\&AS}
\newcommand*\actaa{Acta Astron.}
\newcommand*\aj{AJ}
\newcommand*\an{Astronomische Nachrichten}
\newcommand*\ao{Appl.~Opt.}
\let\applopt\ao
\newcommand*\apj{ApJ}
\newcommand*\apjl{ApJ}
\let\apjlett\apjl
\newcommand*\apjs{ApJS}
\let\apjsupp\apjs
\newcommand*\aplett{Astrophys.~Lett.}
\newcommand*\apspr{Astrophys.~Space~Phys.~Res.}
\newcommand*\apss{Ap\&SS}
\newcommand*\araa{ARA\&A}
\newcommand*\azh{AZh}
\newcommand*\baas{BAAS}
\newcommand*\bac{Bull. astr. Inst. Czechosl.}
\newcommand*\bain{Bull.~Astron.~Inst.~Netherlands}
\newcommand*\caa{Chinese Astron. Astrophys.}
\newcommand*\cjaa{Chinese J. Astron. Astrophys.}
\newcommand*\fcp{Fund.~Cosmic~Phys.}
\newcommand*\gca{Geochim.~Cosmochim.~Acta}
\newcommand*\grl{Geophys.~Res.~Lett.}
\newcommand*\iaucirc{IAU~Circ.}
\newcommand*\icarus{Icarus}
\newcommand*\jcap{J. Cosmology Astropart. Phys.}
\newcommand*\jcp{J.~Chem.~Phys.}
\newcommand*\jgr{J.~Geophys.~Res.}
\newcommand*\jqsrt{J.~Quant.~Spec.~Radiat.~Transf.}
\newcommand*\jrasc{JRASC}
\newcommand*\memras{MmRAS}
\newcommand*\memsai{Mem.~Soc.~Astron.~Italiana}
\newcommand*\mnras{MNRAS}
\newcommand*\na{New A}
\newcommand*\nar{New A Rev.}
\newcommand*\nat{Nature}
\newcommand*\nphysa{Nucl.~Phys.~A}
\newcommand*\pasa{PASA}
\newcommand*\pasj{PASJ}
\newcommand*\pasp{PASP}
\newcommand*\physrep{Phys.~Rep.}
\newcommand*\physscr{Phys.~Scr}
\newcommand*\planss{Planet.~Space~Sci.}
\newcommand*\pra{Phys.~Rev.~A}
\newcommand*\prb{Phys.~Rev.~B}
\newcommand*\prc{Phys.~Rev.~C}
\newcommand*\prd{Phys.~Rev.~D}
\newcommand*\pre{Phys.~Rev.~E}
\newcommand*\prl{Phys.~Rev.~Lett.}
\newcommand*\procspie{Proc.~SPIE}
\newcommand*\qjras{QJRAS}
\newcommand*\rmxaa{Rev. Mexicana Astron. Astrofis.}
\newcommand*\skytel{S\&T}
\newcommand*\solphys{Sol.~Phys.}
\newcommand*\sovast{Soviet~Ast.}
\newcommand*\ssr{Space~Sci.~Rev.}
\newcommand*\zap{ZAp}
\newcommand*\ATel{ATEL}
\newcommand*\NewAR{New Astron. Rev.}
\newcommand*\Obs{Observatory}
\newcommand*\PASA{Publ. Astron. Soc. Aust.}
\newcommand*\PublisherCambridge{(Cambridge: Cambridge University Press)}
\newcommand*\PublisherUAP{Universal Academy Press}



Mauche, C., 2010, http$://cxc.harvard.edu/cdo/accr10/pres/Mauche\_$Chris.pdf	

\label{lastpage}
\end{document}